%% file: main.tex
\documentclass[conference]{IEEEtran}
\IEEEoverridecommandlockouts

\usepackage[cmex10]{amsmath}
\usepackage{graphicx}
\usepackage{amssymb}
\usepackage{amsfonts}
\usepackage[latin1]{inputenc}
\usepackage{color}
\usepackage{cite}
\usepackage{array}
\usepackage{bm}
\usepackage{mathdots}
\usepackage{xcolor}
\usepackage[nonumberlist]{glossaries}
\usepackage{siunitx} 
\usepackage[font=footnotesize]{caption}
\usepackage{subcaption}
\usepackage{microtype} 
\usepackage{relsize}
\usepackage{url}
\usepackage{mwe}
\usepackage{etoolbox}

\def\BibTeX{{\rm B\kern-.05em{\sc i\kern-.025em b}\kern-.08em
    T\kern-.1667em\lower.7ex\hbox{E}\kern-.125emX}}

\input{acronym}

\begin{document}

\title{User Positioning in mmW 5G Networks using Beam-RSRP Measurements and Kalman Filtering}

\author{\IEEEauthorblockN{Elizaveta Rastorgueva-Foi\IEEEauthorrefmark{1}, M\'ario Costa\IEEEauthorrefmark{2}, Mike Koivisto\IEEEauthorrefmark{1}, Kari Lepp\"anen\IEEEauthorrefmark{2}, and Mikko Valkama\IEEEauthorrefmark{1}}

\IEEEauthorblockA{\IEEEauthorrefmark{1} Laboratory of Electronics and Communications Engineering, Tampere University of Technology, Finland}

\IEEEauthorblockA{\IEEEauthorrefmark{2} Huawei Technologies Oy (Finland) Co., Ltd, Finland}
Email: elizaveta.rastorgueva-foi@tut.fi}%

\maketitle

{\let\thefootnote\relax\footnotetext{\footnotesize
This work was supported by the Doctoral Program of the President of Tampere University of Technology, the Tuula and Yrj\"o Neuvo Fund, the Nokia Foundation, and the Finnish Funding Agency for Technology and Innovation (Tekes), under the projects ''TAKE-5: 5th Evolution Take of Wireless Communication Networks", and ''WIVE: Wireless for Verticals".
}}

\begin{abstract}
In this paper, we exploit the 3D-beamforming features of multiantenna equipment employed in \gls{5G} networks, operating in the \gls{mmW} band, for accurate positioning and tracking of users. We consider sequential estimation of users' positions, and propose a two-stage \gls{EKF} that is based on \gls{RSRP} measurements. In particular, beamformed \gls{DL} \glspl{RS} are transmitted by multiple \glspl{BS} and measured by \glspl{UE} employing receive beamforming. The so-obtained \gls{BRSRP} measurements are fed back to the \glspl{BS} where the corresponding \gls{DoDs} are sequentially estimated by a novel \gls{EKF}. Such angle estimates from multiple \glspl{BS} are subsequently fused on a central entity into 3D position estimates of \glspl{UE} by means of an angle-based \gls{EKF}. The proposed positioning scheme is scalable since the computational burden is shared among different network entities, namely \glspl{TRP} and \gls{gNB}, and may be accomplished with the signalling currently specified for 5G. We assess the performance of the proposed algorithm on a realistic outdoor \gls{5G} deployment with a detailed ray tracing propagation model based on the METIS Madrid map. Numerical results with a system operating at $\mathbf{39}$\,\bf{GHz} 
show that sub-meter $3$D positioning accuracy is achievable in future \gls{mmW} \gls{5G} networks.

\end{abstract}

\begin{IEEEkeywords}
5G networks, beamforming, RSRP, positioning, localization, tracking, direction-of-departure, location-awareness, extended Kalman filter, line-of-sight
\end{IEEEkeywords}

\glsresetall

\input{introduction}

\input{system_model}
\input{ekf}
\input{simulations}

\input{conclusion}


\bibliographystyle{IEEEtran/bibtex/IEEEtran}


\end{document}

%% file: acronym.tex
\newacronym{3GPP}{3GPP}{3rd generation partnership project}
\newacronym{5G}{5G}{fifth generation}
\newacronym{5G-NR}{5G-NR}{fifth generation new radio}
\newacronym{ACF}{ACF}{autocorrelation function}
\newacronym{AR}{AR}{auto-regressive}
\newacronym{AN}{AN}{access node}
\newacronym{AoA}{AoA}{angle of arrival}
\newacronym{AWGN}{AWGN}{additive white Gaussian noise}
\newacronym{BB}{BB}{baseband}
\newacronym{BBU}{BBU}{base band unit}
\newacronym{BS}{BS}{base station}
\newacronym{BIC}{BIC}{Bayesian information criterion}
\newacronym{BLER}{BLER}{block error rate}
\newacronym{BPSK}{BPSK}{binary phase shift keying}
\newacronym{CCE}{CCE}{control channel element}
\newacronym{CDF}{CDF}{cumulative distribution function}
\newacronym{CL}{CL}{centroid localization}
\newacronym{CP}{CP}{cyclic prefix}
\newacronym{CPPS}{CPPS}{constant phase \gls{PS}}
\newacronym{CR}{CR}{cognitive radio}
\newacronym{C-RAN}{C-RAN}{centralized radio access network}
\newacronym{CRB}{CRB}{Cram\'er-Rao bound}
\newacronym{CSI}{CSI}{channel state information}
\newacronym{CSIT}{CSIT}{channel state information at transmitter}
\newacronym{CA}{CA}{constant acceleration}
\newacronym{CT}{CT}{constant turn}
\newacronym{CV}{CV}{constant velocity}
\newacronym{D2D}{D2D}{device-to-device}
\newacronym{DBS}{DBS}{different beamwidth sectors}
\newacronym{DCAA}{DCAA}{digitally controlled antenna array}
\newacronym{DCI}{DCI}{downlink control information}
\newacronym{DCPS}{DCPS}{digitally controlled \gls{PS}}
\newacronym{DFU}{DFU}{{}\acrshort{DoA} fusion}
\newacronym{DL}{DL}{downlink}
\newacronym{eMMB}{eMMB}{enhanced mobile broadband}
\newacronym[firstplural=directions of arrival (DoAs)]{DoA}{DoA}{direction of arrival}
\newacronym{DoD}{DoD}{direction of departure}
\newacronym{DoDs}{DoDs}{directions of departure}
\newacronym{gNB}{gNB}{5G-NR Node B}
\newacronym{DS-CDMA}{DC-CDMA}{direct sequence code division multiple access}
\newacronym{MMA}{MMA}{multi-mode antenna}
\newacronym{E2E}{E2E}{end-to-end}
\newacronym{E911}{E911}{enhanced 911}
\newacronym{EADF}{EADF}{effective aperture distribution function}
\newacronym{EBS}{EBS}{equal beamwidth sectors}
\newacronym{EKF}{EKF}{extended Kalman filter}
\newacronym{ESA}{ESA}{equal sector antenna}
\newacronym{ESPAR}{ESPAR}{electronically steerable parasitic array radiator}
\newacronym{EOTD}{E-OTD}{enhanced observed time difference}
\newacronym{EW}{EW}{equal weighting}
\newacronym{FCC}{FCC}{federal communications commission}
\newacronym{FFT}{FFT}{fast Fourier transform}
\newacronym[firstplural=Fisher information matrices (FIM)]{FIM}{FIM}{Fisher information matrix}
\newacronym{GMM}{GMM}{Gaussian mixture model}
\newacronym{GNSS}{GNSS}{global navigation satellite system}
\newacronym{GPS}{GPS}{global positioning system}
\newacronym{GSM}{GSM}{global system for mobile communications}
\newacronym{ICI}{ICI}{inter-carrier-interference}
\newacronym{ICMP}{ICMP}{internet control message protocol}
\newacronym{IoT}{IoT}{Internet-of-Things}
\newacronym{ISD}{ISD}{inter-site distance}
\newacronym{ISI}{ISI}{inter-symbol-interference}
\newacronym{ITS}{ITS}{intelligent transportation system}
\newacronym{ITU}{ITU}{international telecommunication union}
\newacronym{KF}{KF}{Kalman filter}
\newacronym{LDPC}{LDPC}{low-density parity-check}
\newacronym{LMMSE}{LMMSE}{linear minimum mean-square error}
\newacronym{LoS}{LoS}{line-of-sight}
\newacronym{LS}{LS}{least squares}
\newacronym{TLS}{TLS}{total least squares}
\newacronym{NLLS}{NLLS}{nonlinear least squares}
\newacronym{LTE}{LTE}{long term evolution}
\newacronym{LWA}{LWA}{leaky-wave antenna}
\newacronym{MAP}{MAP}{maximum a posteriori}
\newacronym{MCS}{MCS}{modulation and coding scheme} 
\newacronym{METIS}{METIS}{Mobile and wireless communications enablers for the twenty-twenty information society}
\newacronym{MF}{MF}{matched filter}
\newacronym{MGSCM}{MGSCM}{METIS geometry-based stochastic channel model}
\newacronym{MIMO}{MIMO}{multiple-input multiple-output}
\newacronym{MSE}{MSE}{mean-squared error}
\newacronym{ML}{ML}{maximum likelihood}
\newacronym{MLE}{MLE}{maximum likelihood estimator}
\newacronym{MMSE}{MMSE}{minimum mean-square error}
\newacronym{MTC}{MTC}{machine-type communication}
\newacronym{mMTC}{mMTC}{massive \gls{MTC}}
\newacronym{mmW}{mmW}{millimeter wave}
\newacronym{MU-MIMO}{MU-MIMO}{multiuser multiple-input-multiple-output}
\newacronym{MU-MISO}{MU-MISO}{multiuser multiple-input-single-output}
\newacronym{MVUE}{MVUE}{minimum variance unbiased estimator}
\newacronym{NLoS}{NLoS}{non-line-of-sight}
\newacronym{NR}{NR}{new radio}
\newacronym{ppm}{ppm}{parts per million}
\newacronym{OFDM}{OFDM}{orthogonal frequency-division multiplexing}
\newacronym{OFDMA}{OFDMA}{orthogonal frequency-division multiple access}
\newacronym{ON}{ON}{observing node}
\newacronym{OTDoA}{OTDoA}{observed time difference of arrival}
\newacronym{PDF}{PDF}{probability density function}
\newacronym{PDoA}{PDoA}{phase difference of arrival}
\newacronym{PDCCH}{PDCCH}{physical downlink control channel}
\newacronym{QPSK}{QPSK}{quadrature phase shift keying}
\newacronym{NR-PDCCH}{NR-PDCCH}{\acrshort{NR} \acrlong{PDCCH}}
\newacronym{REG}{REG}{resource element group}
\newacronym{REM}{REM}{radio environment map}
\newacronym{RAT}{RAT}{radio access technology}
\newacronym{RToA}{RToA}{round-trip time of arrival}
\newacronym{RZF}{RZF}{regularized zero-forcing}
\newacronym{PPMCC}{PPMCC}{Pearson product-moment correlation coefficient}
\newacronym{PU}{PU}{primary user}
\newacronym{PS}{PS}{phase shifter}
\newacronym{PW}{PW}{power weighting}
\newacronym{RMSE}{RMSE}{root-mean-squared error}
\newacronym{RRMSE}{RRMSE}{relative root-mean-squared error}
\newacronym{RS}{RS}{reference signal}
\newacronym{RSS}{RSS}{received signal strength}
\newacronym{RRH}{RRH}{remote radio head}
\newacronym{RRM}{RRM}{radio resource management}
\newacronym{RF}{RF}{radio frequency}
\newacronym{RFI}{RFI}{radio frequency interference}
\newacronym{Rx}{Rx}{receiver}
\newacronym{SBS}{SBS}{switched-beam system}
\newacronym{SDE}{SDE}{sector-pair \acrshort{DoA} estimation}
\newacronym{SIMO}{SIMO}{single-input-multiple-output}
\newacronym{SINR}{SINR}{signal-to-interference-plus-noise ratio}
\newacronym{SLS}{SLS}{simplified least squares}
\newacronym{SNR}{SNR}{signal-to-noise ratio}
\newacronym{SSP}{SSP}{side-sector suppression}
\newacronym{SSL}{SSL}{sector selection}
\newacronym{STD}{STD}{standard deviation}
\newacronym{SU}{SU}{secondary user}
\newacronym{SZF}{SZF}{suppressed zero-forcing}
\newacronym{TCP}{TCP}{transmission control protocol}
\newacronym{TDD}{TDD}{time division duplex}
\newacronym{TDoA}{TDoA}{time difference of arrival}
\newacronym{TN}{TN}{target node}
\newacronym[firstplural=times of arrival (ToA)]{ToA}{ToA}{time of arrival}
\newacronym{ToF}{ToF}{time of flight}
\newacronym{TRP}{TRP}{transmission/reception point}
\newacronym{TSLS}{TSLS}{three-stage SLS}
\newacronym{TTI}{TTI}{transmission time interval}
\newacronym{Tx}{Tx}{transmitter}
\newacronym{UCA}{UCA}{uniform circular array}
\newacronym{UAV}{UAV}{unmanned aerial vehicle}
\newacronym{UDN}{UDN}{ultra-dense network}
\newacronym{UE}{UE}{user equipment}
\newacronym{UKF}{UKF}{unscented Kalman filter}
\newacronym{UL}{UL}{uplink}
\newacronym{ULA}{ULA}{uniform linear array}
\newacronym{URLLC}{URLLC}{ultra-reliable low latency communication}
\newacronym{UTDoA}{U-TDoA}{uplink-time difference of arrival}
\newacronym{UTD}{UTD}{the uniform theory of diffraction}
\newacronym{UMi}{UMi}{urban micro}
\newacronym{UN}{UN}{user node}
\newacronym{VW}{VW}{variance weighting}
\newacronym{WCL}{WCL}{weighted centroid localization}
\newacronym{WLAN}{WLAN}{wireless local area network}
\newacronym{wrt}{wrt}{with respect to}
\newacronym{WSS}{WSS}{wide sense stationary}
\newacronym{ZF}{ZF}{zero-forcing}
\newacronym{BRSRP}{BRSRP}{beam-\gls{RSRP}}
\newacronym{RSRP}{RSRP}{reference signal received power}
\newacronym{pdf}{pdf}{probability density function}
\newacronym{PF}{PF}{particle filter}
\newacronym{NLLF}{NLLF}{negative log likelihood function}
\newacronym{ARU}{ARU}{analog radio unit}
\newacronym{DF}{DF}{direction finding}

%% file: introduction.tex
\section{Introduction} \label{sec:intro}
The adoption of the \gls{mmW} frequency bands by the \gls{5G} wireless networks allows not only for a tremendous increase in capacity but also opens new opportunities for high-accuracy \gls{UE} positioning. In fact, in 3GPP, a study item proposal on 5G positioning using \gls{RAT}-dependent solutions is currently under discussion \cite{3GPPTR22872,3GPPTP172746}. In particular, \gls{5G} \glspl{BS} and \glspl{UE} operating at \gls{mmW} frequencies are expected to make a considerable use of transmit and receive beamforming due to path-loss at such frequencies \cite{SRHNR14,3GPPTS38214}. In addition to improved resource utilization, beamforming at \glspl{BS} can be used for estimating the \gls{DoD} of a \gls{DL} signal, which in turn can be exploited for high-accuracy positioning of a \gls{UE}. 

In this paper, we propose a sequential estimation method for user positioning based on the  beamformed \gls{DL} \gls{RSRP} measurements from a given \gls{UE}. Such \gls{BRSRP} measurements are employed by a novel two-stage \gls{EKF} for estimating and tracking the 3D position of the \glspl{UE}. In particular, each \gls{BS} transmits beamformed \gls{DL} \glspl{RS} that are measured by the \glspl{UE} employing receive beamforming. The measured \gls{BRSRP} values are then communicated back to the \glspl{BS} where the corresponding \glspl{DoD} are sequentially estimated by the first stage \gls{EKF}. Thereafter, in the second \gls{EKF} stage, the \gls{UE} specific \gls{DoD} estimates from the previous stage \glspl{EKF} from all the available \glspl{BS} are fused in order to obtain the sequential 3D position estimates for a given \gls{UE}. 

\gls{BRSRP} measurements make it possible for the proposed algorithm to be deployed on analog beamforming architectures, which are known to be significantly less expensive than fully-digital or even hybrid architectures, and thus more suitable for \gls{mmW} applications. Moreover, exploiting feedback of \gls{DL} \gls{RS} measurements allows our \gls{EKF} to be directly applicable to \gls{5G} networks, and therefore provide highly accurate $3$D positioning  of users with essentially the currently agreed specification for $5$G \cite{3GPPTS38214}. The \gls{EKF} algorithm typically outperforms batch estimation schemes, and provides a good trade-off between performance and complexity when compared to other sequential estimation techniques such as particle-filtering. The main advantage of the cascaded two-stage scheme considered herein is that the computational load can be distributed between \glspl{BS} and a central entity, which also leads to a reduction on the signalling load while the central entity is tracking the \gls{UE}.


 This work can be understood as an extension of the work in \cite{KCWH17} to the case of \gls{BRSRP} measurements, instead of using the relative phases of the \gls{UL} signals received across \gls{BS}'s antennas for \glspl{UE} positioning. Recent applications of \gls{BRSRP} measurements to angle estimation include \cite{PZJD17}, where \gls{RSRP} measurements are carried out with a single \gls{MMA}. In a case of \glspl{MMA}, directional power measurements are enabled by the registration of the antenna surface current distribution corresponding to the different characteristic modes. Also, in \cite{LBS10} \gls{DoA} estimation via single-antenna \gls{RSRP} measurements by exploiting the antenna radiation pattern diversity is proposed. Unlike this paper, the work in \cite{PZJD17,LBS10} focused on batch techniques for \gls{DoA} estimation, and did not consider user positioning. In fact, sequential estimation typically outperforms batch schemes due to the ability to fuse measurements from consecutive time-instants \cite{Kay93}, thus making it suitable for tracking moving users.


The rest of the paper is organized as follows. First, the considered system model is introduced and described in Section~\ref{sec:system_model}. Both stages of the proposed \gls{EKF} solution, i.e., the \gls{DoD} tracking and \gls{UE} positioning \glspl{EKF}, are derived and explained in detailed manner in Section~\ref{sec:ekf}. Thereafter, the considered simulation scenarios as well as the results of our simulations and numerical evaluations are presented in Section~\ref{sec:simulations}. Finally, Section~\ref{sec:conclusions} concludes the paper.

%% file: system_model.tex
\section{System Model} \label{sec:system_model}

Let $\bm{y}_{i,j} \in \mathbb{C}^{\mathcal{M}_f}$ denote the multicarrier observation in an \gls{OFDM} system at the \gls{UE} side. The subscripts $i,j$ refer to the $i$th \gls{UE} \gls{Rx} beam and the $j$th \gls{BS} \gls{Tx} beam, and $\mathcal{M}_f$ denotes the number of subcarriers. Assuming a single dominant propagation path, the observation at the \gls{UE} is given by
\begin{equation} \label{eq:freq_dom_out}
    \bm{y}_{i,j} = \bm{S} \bm{b}_f \bm{b}_{\mathrm{UE}}^{i\,{\textrm{T}}}(\vartheta_a,\varphi_a) \bm{\Gamma} \bm{b}_{\mathrm{BS}}^j(\vartheta_d,\varphi_d) + \bm{n}_{i,j},
\end{equation}
where $\bm{S} \in \mathbb{C}^{\mathcal{M}_f \times \mathcal{M}_f}$ is a diagonal matrix denoting the transmitted symbols in frequency domain, and $\bm{b}_f \in \mathbb{C}^{\mathcal{M}_f}$ denotes the combined frequency-response of the channel and \gls{Tx}-\gls{Rx} \gls{RF}-chains. Moreover, $\bm{b}_{\mathrm{BS}}^j(\vartheta_d,\varphi_d) \in \mathbb{C}^{2}$ and $\bm{b}_{\mathrm{UE}}^i(\vartheta_a,\varphi_a) \in \mathbb{C}^{2}$ denote the complex-valued polarimetric beampattern of the $j$th \gls{BS} and $i$th \gls{UE} beams, respectively. Here, the departure elevation and azimuth angles at the \gls{BS} are denoted as $(\vartheta_d,\varphi_d)$, whereas the arrival elevation and azimuth angles at the \gls{UE} are denoted as $(\vartheta_a,\varphi_a)$. Finally, $\bm{\Gamma} \in \mathbb{C}^{2\times2}$ denotes the channel's polarimetric path-weights \cite{Mol04,Ric05}, and $\bm{n}_{i,j} \in \mathbb{C}^{\mathcal{M}_f}$ denotes measurement noise. In particular, we assume that $\bm{n}_{i,j} \sim \mathcal{N}_C(\bm{0},\tilde{\sigma}^2_{i,j} \bm{I})$, $\tilde{\sigma}^2_{i,j} = \tilde{\sigma}^2~\forall (i,j)$, as well as $\mathbb{E}\{\bm{n}_{i,j} \bm{n}_{k,l}^H\} = \bm{0}$ when $i\neq k$. In other words, we assume a noise-limited system and a radio channel with negligible diffuse scattering. The assumption of the uncorrelated measurement noise holds when the \gls{UE} beams are formed at different time-instants, employ different \gls{RF}-chains, or the beams are orthogonal. These assumptions typically hold in \gls{mmW} systems.

For polarimetric beampatterns we have \begin{equation}
\bm{b}_{\mathrm{BS}}^j(\vartheta_d,\varphi_d) = [b_{\mathrm{BS}_\theta}^j(\vartheta_d,\varphi_d), b_{\mathrm{BS}_\phi}^j(\vartheta_d,\varphi_d)]^{\textrm{T}},
\end{equation}
where the subscripts $\theta$ and $\phi$ denote the orthogonal components, along the tangential spherical unit-vectors, of the electric-field corresponding to the $j$th \gls{BS} beam. A similar representation is considered also for $\bm{b}_{\mathrm{UE}}^i(\vartheta_a,\varphi_a)$.

We now proceed by considering two limitations commonly found in practice. Firstly, the \gls{UE}'s \gls{Rx} beam characteristics are either not available at the network side or the capacity of the feedback channel does not allow for reporting all $\mathcal{M}_{\mathrm{BS}} \times \mathcal{M}_\mathrm{UE}$ channels, where $\mathcal{M}_{\mathrm{BS}}$ and $\mathcal{M}_{\mathrm{UE}}$ denote the number of beams at a given \gls{BS} and \gls{UE}, respectively. Hence, we focus on estimating the \gls{DoD} of the \gls{DL} \gls{LoS} path. Note that both \gls{DoD} and \gls{DoA} may be estimated given that all of the $\mathcal{M}_{\mathrm{BS}} \times \mathcal{M}_\mathrm{UE}$ channels are available at the \gls{BS}. Secondly, the relative phases among the \gls{BS} \gls{Tx} beams are unknown (e.g., uncalibrated system). Hence, \gls{RSRP} measurements of the \gls{BS} \gls{Tx} beams, that are robust against the above limitations, are used for estimating the position of the \gls{UE}. In particular, we consider \gls{BRSRP} measurements defined as
\begin{equation}
    \beta_{i,j} = \frac{1}{\mathcal{M}_f} \sum_{m=1}^{\mathcal{M}_f} |[\bm{y}_{i,j}]_m|^2.
\end{equation}
Note that this is similar to \gls{RSRP} measurements commonly used in wireless communication systems.

\begin{figure}[!t]
\centering
\includegraphics[width=\linewidth]{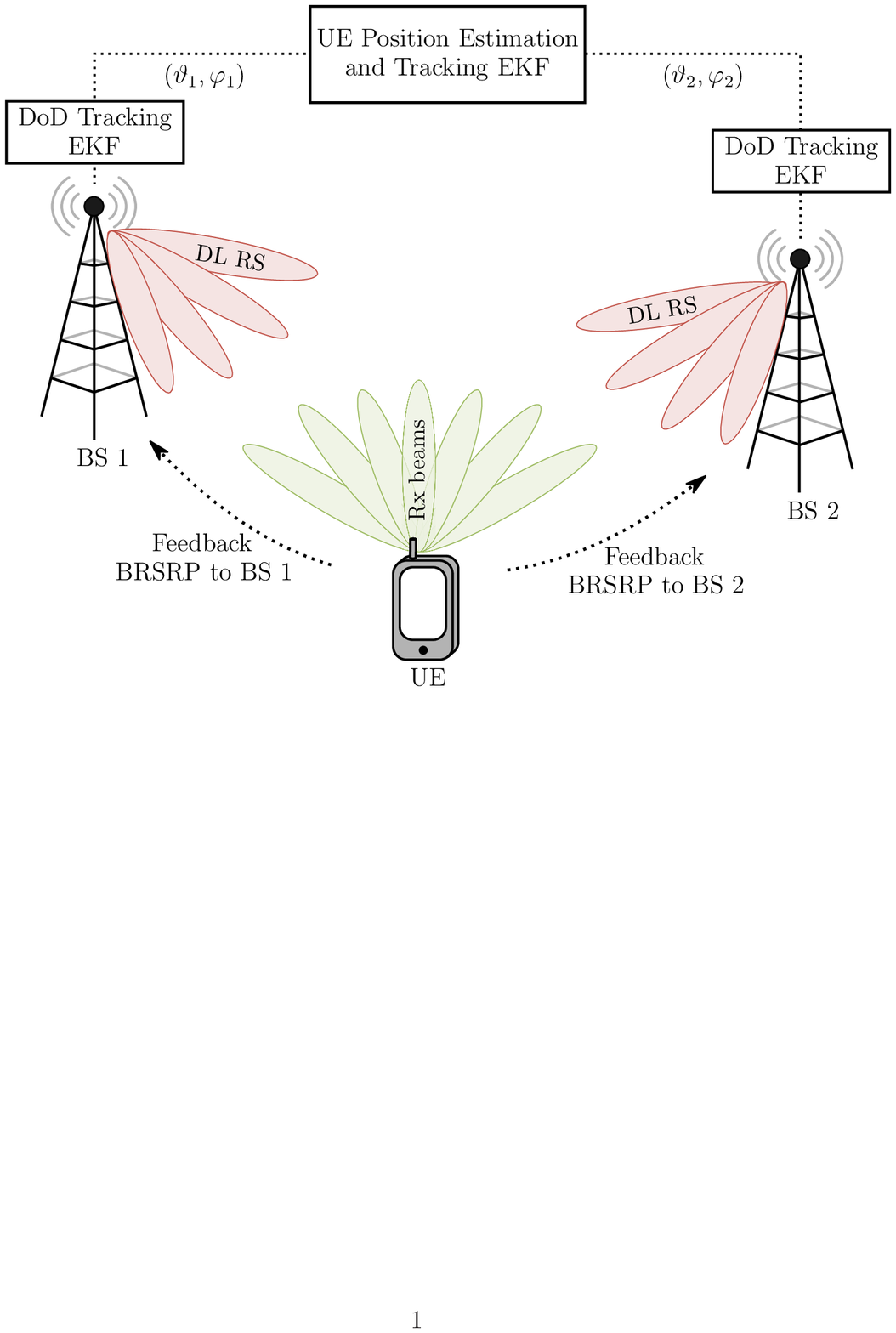}
\caption{Illustration of the \gls{UE} positioning approach considered in this paper. In particular, \glspl{UE} feedback \gls{RSRP} measurements obtained from \gls{DL} \glspl{RS} and transmitted across multiple \gls{BS} beams. Only angles are used for positioning purposes. A single \gls{BS} suffices in determining the $2$D position of a \gls{UE} given that its height is known. For $3$D positioning at least two \glspl{BS} are needed.}
\label{fig:sol_sch}
\end{figure}

The problem addressed in this paper is that of sequentially estimating the \gls{UE} position based on feedback \gls{BRSRP} measurements as illustrated in Fig.~\ref{fig:sol_sch}.

%% file: ekf.tex
\section{Proposed Extended Kalman Filter} \label{sec:ekf}

We now consider sequential estimation of the \gls{UE}'s $3$D position by means of a two-stage \gls{EKF}. In particular, each \gls{BS} employs an \gls{EKF} for estimating and tracking the \gls{DoD} using feedback \gls{BRSRP} measurements from the \gls{UE}. This is the first stage of the sequential estimation procedure. The second stage \gls{EKF} consists in fusing the \glspl{DoD} according to their covariance matrices, both tracked by the first stage \glspl{EKF}, into position estimates. We follow the so-called information form \gls{EKF} instead of the more widely used Kalman-gain formulation. In fact, the former is computationally more attractive than the latter when the state-vector has smaller dimension than the observation (or measurement) vector. For example, for $\mathcal{M}_\mathrm{BS}=64$ beams and $2$ parameters composing the state-vector ($\vartheta, \varphi$) the information-form \gls{EKF} needs to invert a $2\times2$ matrix while in the Kalman-gain formulation a $64\times64$ matrix inversion is required at each step. Note that the two-stage \gls{EKF} proposed in this section may be understood as an extension of the work in \cite{KCWH17} to \gls{BRSRP} measurements, instead of using relative-phase measurements.

\subsection{EKF for DoD Estimation and Tracking}

The state vector for the \gls{DoD}-\gls{EKF} is $\bm{s} = [\vartheta, \varphi, \Delta \vartheta, \Delta \varphi]^{\textrm{T}}$, where $\Delta \vartheta$ and $\Delta \varphi$ denote the rate-of-change of $\vartheta$ and $\varphi$, respectively. The prediction step of the \gls{DoD}-\gls{EKF} is then
\begin{align}
    \bm{s}^-[n] & = \bm{F} \bm{s}^+[n-1] \\
    \bm{C}^-[n] & = \bm{F} \bm{C}^+[n-1] \bm{F}^{\textrm{T}} + \bm{Q},
\end{align}
where $\bm{F} \in \mathbb{R}^{4\times4}$, $\bm{C}\in\mathbb{R}^{4\times4}$, and $\bm{Q}\in\mathbb{R}^{4\times4}$ denote the state-transition matrix, state covariance matrix, and state-noise covariance matrix, respectively. Matrices $\bm{F}$ and $\bm{Q}$ can be found from \cite[Ch.2]{HSS11} by noting that we have employed a continuous white-noise acceleration model for the state-dynamics. The update step of the \gls{DoD}-\gls{EKF} is
\begin{align} \label{eq:ekf_dod}
    \bm{C}^+[n] & = \left(\bm{C}^-[n]^{-1} + \bm{\mathcal{I}}(\bm{s}^-[n])\right)^{-1} \\
    \Delta \bm{s}[n] & = \bm{C}^+[n] \, \bm{q}(\bm{s}^-[n]) \\
    \bm{s}^+[n] & = \bm{s}^-[n] + \Delta \bm{s}[n],
\end{align}
where $\bm{\mathcal{I}}(\bm{s}^-[n]) \in \mathbb{R}^{4 \times 4}$
and $\bm{q}(\bm{s}^-[n])\in\mathbb{R}^{4}$ denote the observed \gls{FIM} and gradient of the log-likelihood function of the state
given \gls{BRSRP} measurements, respectively.

In particular, \gls{BRSRP} measurements can be shown to follow a noncentral $\chi^2$-distribution with a \gls{pdf} given by \cite{PZJD17},\cite[Ch.2]{PS08}
\begin{align} \begin{split}\label{eq:chi_squared}
p(\beta_{i,j}) = &\frac{\mathcal{M}_f}{\tilde{\sigma}_{i,j}^2}\left(\frac{\mathcal{M}_f\beta_{i,j}}{\lambda_{i,j}}\right)^{\frac{\mathcal{M}_f-1}{2}} \textrm{e}^{-\frac{\lambda_{i,j}+\mathcal{M}_f\beta_{i,j}}{\tilde{\sigma}_{i,j}^2}}\\ & \times I_{\mathcal{M}_f-1}\left(\frac{2\sqrt{\lambda_{i,j}\mathcal{M}_f\beta_{i,j}}}{\tilde{\sigma}_{i,j}^2}\right),
\end{split}
\end{align}
where $I_x(\cdot) \in \mathbb{R}$ denotes a modified Bessel function of the first kind. For a growing number of subcarriers $\mathcal{M}_f$, the \gls{pdf} in \eqref{eq:chi_squared} approaches a Gaussian \cite{PZJD17}, and we thus have 
\begin{equation}
    \beta_{i,j} \sim \mathcal{N}(\mu_{i,j},\sigma_{i,j}^2).
\end{equation}
Here, the mean and variance of \gls{BRSRP} measurements are given, respectively, by
\begin{align}
    \mu_{i,j} & = \frac{\lambda_{i,j}}{\mathcal{M}_f} + \tilde{\sigma}_{i,j}^2 \\ \label{eq:brsrp_var}
    \sigma_{i,j}^2 & =  \frac{\tilde{\sigma}_{i,j}^4}{\mathcal{M}_f} + \frac{2\tilde{\sigma}_{i,j}^2 \lambda_{i,j}}{\mathcal{M}_f^2},
\end{align}
where 
\begin{align}
    \lambda_{i,j} & = \sum_{m=1}^{\mathcal{M}_f} |[\bm{S}]_m [\bm{b}_f]_m \gamma_{i,j}|^2 \\
    \gamma_{i,j} & = \bm{b}_{\mathrm{UE}}^{i \,\textrm{T}}(\vartheta_a,\varphi_a) \bm{\Gamma} \bm{b}_{\mathrm{BS}}^j(\vartheta_d,\varphi_d).
\end{align}

Let $\bm{\beta}\in\mathbb{R}^{\mathcal{M}_\mathrm{BS}}$ denote the \gls{BRSRP} measurements for all \gls{BS} beams and a given \gls{UE} beam. The chosen \gls{UE} beam can be the one that yields the largest sum of \gls{BS}'s \gls{BRSRP} measurements among all \gls{UE} beams, i.e. $\max_i \sum_{j=1}^{\mathcal{M}_\mathrm{BS}} \beta_{i,j}$, or simply the \gls{UE} beam corresponding to the largest \gls{BRSRP} measurement, for example. It follows that $\bm{\beta} \sim \mathcal{N}(\bm{\mu}(\bm{\Theta}),\bm{C}(\bm{\Theta}))$, where
\begin{align} \label{eq:gaussian_approx_brsrp}
\bm{\mu}(\bm{\Theta}) & = \bm{A}(\vartheta_d,\varphi_d) \bm{\alpha}\frac{P_{\mathrm{Tx}}}{\mathcal{M}_f} + \bm{1} \tilde{\sigma}^2 \\ \bm{C}(\bm{\Theta}) & = \mathrm{diag}\left\{\bm{A}(\vartheta_d,\varphi_d) \bm{\alpha} \frac{2\tilde{\sigma}^2P_{\mathrm{Tx}}}{\mathcal{M}_f^2} + \bm{1} \frac{\tilde{\sigma}^4}{\mathcal{M}_f} \right\}.
\end{align}
Here, $\bm{\Theta}=[\vartheta_d,\varphi_d,\alpha_1,\alpha_4,\Re\{\alpha_2\},\Im\{\alpha_2\},\tilde{\sigma}^2]^{\textrm{T}}$ denotes the unknown parameter vector and $\bm{\alpha} = [\alpha_1,\ldots,\alpha_4]^{\textrm{T}}$ is given by $\bm{\alpha} = \mathrm{vec}\{\bm{\Gamma}^H \bm{b}^{i^\ast}_{\mathrm{UE}}(\vartheta_a,\varphi_a) \bm{b}^{i\,{\textrm{T}}}_{\mathrm{UE}}(\vartheta_a,\varphi_a)\bm{\Gamma}\}$. Moreover, $P_{\mathrm{Tx}} = \sum_{m=1}^{\mathcal{M}_f} |[\bm{S}]_m [\bm{b}_f]_m|^2$ and $\bm{A}(\vartheta_d,\varphi_d)\in\mathbb{C}^{\mathcal{M}_{\mathrm{BS}} \times 4}$ is given by
\begin{align}\begin{split}
\bm{A}(\vartheta_d,\varphi_d) = [&\bm{b}^\ast_{\mathrm{BS}_\phi}(\vartheta_d,\varphi_d) \odot \bm{b}_{\mathrm{BS}_\phi}(\vartheta_d,\varphi_d),\\& \bm{b}^\ast_{\mathrm{BS}_\theta}(\vartheta_d,\varphi_d) \odot \bm{b}_{\mathrm{BS}_\phi}(\vartheta_d,\varphi_d), \\ & \bm{b}^\ast_{\mathrm{BS}_\phi}(\vartheta_d,\varphi_d) \odot \bm{b}_{\mathrm{BS}_\theta}(\vartheta_d,\varphi_d),\\& \bm{b}^\ast_{\mathrm{BS}_\theta}(\vartheta_d,\varphi_d) \odot \bm{b}_{\mathrm{BS}_\theta}(\vartheta_d,\varphi_d)],
\end{split}
\end{align}
where $\odot$ denotes the Hadamard-Schur (element-wise) product. 

The \gls{EKF} can now be implemented with the observed \gls{FIM} and gradient of the log-likelihood function of $\bm{\Theta} \in \mathbb{R}^7$ by exploiting the asymptotic (Gaussian) distribution of \gls{BRSRP} measurements. Convenient expressions for the \gls{FIM} and gradient under Gaussian distributed observations can be found in \cite[Ch.3]{Kay93}. However, such an approach may not be computationally attractive since one would need to track $7$ parameters out of which only $2$ are of interest for angle based positioning. We would thus need to track $5$ nuisance parameters. Since the computational complexity of each \gls{EKF} iteration is typically $\mathcal{O}(n^3)$, where $n$ denotes the dimension of the state-vector, it is important in practice to formulate the sequential estimation problem at hand in a way that only the \gls{DoD} is tracked at each \gls{BS}.

Let us thus define the received \gls{SNR} of the $(i,j)$ beam pair as
\begin{equation}
    \mathrm{SNR}_{i,j} \triangleq \frac{\lambda_{i,j}}{\mathcal{M}_f \tilde{\sigma}^2_{i,j}}.
\end{equation}
In the low \gls{SNR} regime we have $\mathcal{M}_f \tilde{\sigma}^2_{i,j} > \lambda_{i,j}$. Moreover, the break-even point between both terms composing the variance of \gls{BRSRP} measurements in \eqref{eq:brsrp_var} is $\mathcal{M}_f \tilde{\sigma}^2_{i,j} = 2 \lambda_{i,j}$. Hence, we make a low-\gls{SNR} approximation of the covariance of $\bm{\beta}$ and assume it is independent of the \gls{DoD}. The resulting log-likelihood function is
\begin{align} \label{eq:log-likelihood_approx}
\begin{split}
    \ell_\mathrm{aprx}(\bm{\Theta},\sigma^2|\bm{\beta})  = &-\frac{\mathcal{M}_\mathrm{BS}}{2} \ln{2\pi} -\frac{\mathcal{M}_\mathrm{BS}}{2} \ln{\sigma^2}\\ 
    & - \frac{1}{2\sigma^2}\|\bm{\beta}-\bm{\mu}(\bm{\Theta}))\|^2.
\end{split}
\end{align}
The above log-likelihood function is separable in $(\vartheta_d,\varphi_d)$ and $(\bm{\alpha},\tilde{\sigma}^2,\sigma^2)$ since the \gls{MLE} of the latter parameters may be found in a closed-form for a given $(\vartheta,\varphi)$ \cite{GP03,OVK92}. Hence, the concentrated log-likelihood function is
\begin{align}
\begin{split}
    \label{eq:log-likelihood_approx_conc}
    \ell_\mathrm{caprx}(\vartheta_d,\varphi_d|\bm{\beta}) = & -\frac{\mathcal{M}_{\mathrm{BS}}}{2} \ln 2\pi - \frac{\mathcal{M}_{\mathrm{BS}}}{2} \\
    &- \frac{\mathcal{M}_{\mathrm{BS}}}{2} \ln \frac{\|\bm{P}^\bot_{A 1}(\vartheta_d,\varphi_d) \bm{\beta}\|^2}{\mathcal{M}_{\mathrm{BS}}}.
\end{split}
\end{align}
Such an expression is obtained by replacing $\bm{\alpha}$ and $\sigma^2$ in \eqref{eq:log-likelihood_approx} with the corresponding \glspl{MLE}:
\begin{align}
\begin{split}
        \begin{bmatrix} \hat{\bm{\alpha}}\\\hat{\tilde{\sigma}}^2
    \end{bmatrix} & = \left[\bm{A}(\vartheta_d,\varphi_d) \frac{P_{\mathrm{Tx}}}{\mathcal{M}_f}, \, \bm{1}\right]^\dag \bm{\beta} \\
    \hat{\sigma}^2 & = \frac{1}{\mathcal{M}_{\mathrm{BS}}}\|\bm{P}^\bot_{A 1}(\vartheta_d,\varphi_d) \bm{\beta}\|^2.
\end{split}
\end{align}
Here, $(\cdot)^\dag$ denotes the Moore-Penrose pseudo-inverse. Moreover, $\bm{P}^\bot_{A 1}(\vartheta_d,\varphi_d)=\bm{I}-\bm{P}_{A 1}(\vartheta_d,\varphi_d)$, and $\bm{P}_{A 1}(\vartheta_d,\varphi_d)\in\mathbb{R}^{\mathcal{M}_\mathrm{BS}\times\mathcal{M}_\mathrm{BS}}$ denotes an orthogonal projection matrix. We note that $\bm{P}_{A 1}(\vartheta_d,\varphi_d) = \bm{\mathcal{P}}_{A}(\vartheta_d,\varphi_d) + \bm{\mathcal{P}}_{1}$, where $\bm{\mathcal{P}}_{A}(\vartheta_d,\varphi_d)$ and $\bm{\mathcal{P}}_{1}$ denote oblique projection matrices. In particular, the range-space of $\bm{\mathcal{P}}_{A}(\vartheta_d,\varphi_d)$ is spanned by the columns of $\bm{A}(\vartheta_d,\varphi_d)$ while its nullspace contains a subspace spanned by vector $\bm{1}$. Similarly, the range-space of $\bm{\mathcal{P}}_{1}$ is spanned by vector $\bm{1}$ while its nullspace contains a subspace spanned by the columns of $\bm{A}(\vartheta_d,\varphi_d)$ \cite{BS94}.

The gradient and observed \gls{FIM} (or Hessian) of the concentrated log-likelihood function now follows from the results in \cite{SN89,VOK91}\footnote{To be precise, we have employed $\mathrm{exp}\{\ell_\mathrm{caprx}(\vartheta_d,\varphi_d|\bm{\beta})\}$. Such an operation does not change the global maximum of the log-likelihood function.}
\begin{align}
    [\bm{q}(\vartheta_d,\varphi_d)]_1 & = 2\left(\frac{\partial}{\partial \vartheta_d}\bm{P}^\bot_{A 1}(\vartheta_d,\varphi_d) \, \bm{\beta}\right)^{\textrm{T}} \bm{P}^\bot_{A 1}(\vartheta_d,\varphi_d)\, \bm{\beta} \\
    [\bm{q}(\vartheta_d,\varphi_d)]_2 & = 2\left(\frac{\partial}{\partial \varphi_d}\bm{P}^\bot_{A 1}(\vartheta_d,\varphi_d) \, \bm{\beta}\right)^{\textrm{T}} \bm{P}^\bot_{A 1}(\vartheta_d,\varphi_d)\, \bm{\beta} \\
    [\bm{\mathcal{I}}(\vartheta_d,\varphi_d)]_{1,2} & \approx 2 \left(\frac{\partial}{\partial \vartheta_d}\bm{P}^\bot_{A 1}(\vartheta_d,\varphi_d) \, \bm{\beta}\right)^{\textrm{T}} \nonumber \\
    & \hspace{2cm} \times \frac{\partial}{\partial \varphi_d}\bm{P}^\bot_{A 1}(\vartheta_d,\varphi_d) \, \bm{\beta}.
\end{align}
Note that we have used a first-order approximation of the observed \gls{FIM} since it is known to provide improved convergence \cite{VOK91}. The proposed \gls{EKF} for \gls{DoD} employs the above gradient and observed \gls{FIM} in the update-step. Next, the \glspl{DoD} and corresponding covariance matrices tracked by the \gls{DoD}-\gls{EKF} are used for tracking the \gls{UE} position.

\subsection{EKF for UE Positioning}

The \gls{DoD} estimates tracked by the \gls{EKF} proposed in the previous section can be assumed to be given by
\begin{equation} \label{eq:dod_dist}
\begin{bmatrix}
    \hat{\vartheta}_{k}\\ \hat{\varphi}_k
    \end{bmatrix} \sim \mathcal{N}\left(\begin{bmatrix}
    \vartheta_k\\ \varphi_k
\end{bmatrix},\bm{C}_k\right),
\end{equation}
where the subscript $k$ denotes the \gls{BS} index. Note that the covariance $\bm{C}_k\in\mathbb{R}^{2\times2}$ equals the upper-left $(2\times2)$ block of $\bm{C}^+[n]$ in the \gls{DoD}-\gls{EKF}, and it is assumed to be angle-independent. Such a simplifying assumption is taken here since a closed-form expression for $\bm{C}_k$ is typically rather involved, which in turn would significantly increase the complexity of the Pos-\gls{EKF} proposed in this section. In particular, let the state vector be given by $\bm{s}_\mathrm{UE} = [x_\mathrm{UE}, y_\mathrm{UE}, z_\mathrm{UE}, v_x, v_y, v_z]^{\textrm{T}}$. The prediction step of the Pos-\gls{EKF} is then
\begin{align}
    \bm{s}_\mathrm{UE}^-[n] & = \bm{F}_\mathrm{UE} \bm{s}_\mathrm{UE}^+[n-1] \\
    \bm{C}^-_\mathrm{UE}[n] & = \bm{F}_\mathrm{UE} \bm{C}_\mathrm{UE}^+[n-1] \bm{F}^{\textrm{T}}_\mathrm{UE} + \bm{Q}_\mathrm{UE},
\end{align}
where $\bm{F}_\mathrm{UE} \in \mathbb{R}^{6\times6}$, $\bm{C}_\mathrm{UE}\in\mathbb{R}^{6\times6}$, and $\bm{Q}_\mathrm{UE}\in\mathbb{R}^{6\times6}$ denote the state-transition matrix, state covariance matrix, and state-noise covariance matrix, respectively. Similarly to the \gls{DoD}-\gls{EKF}, matrices $\bm{F}_\mathrm{UE}$ and $\bm{Q}_\mathrm{UE}$ can be found from \cite[Ch.2]{HSS11} by noting that we have employed a continuous white-noise acceleration model for the \gls{UE} state-dynamics. The update step of the Pos-\gls{EKF} is
\begin{align} \label{eq:ekf_pos}
    \bm{C}^+_\mathrm{UE}[n] & = \left(\bm{C}^-_\mathrm{UE}[n]^{-1} + \bm{\mathcal{I}}_\mathrm{UE}(\bm{s}^-_\mathrm{UE}[n])\right)^{-1} \\
    \Delta \bm{s}_\mathrm{UE}[n] & = \bm{C}^+_\mathrm{UE}[n] \, \bm{q}_\mathrm{UE}(\bm{s}^-_\mathrm{UE}[n]) \\
    \bm{s}^+_\mathrm{UE}[n] & = \bm{s}^-_\mathrm{UE}[n] + \Delta \bm{s}_\mathrm{UE}[n],
\end{align}
where $\bm{\mathcal{I}}_\mathrm{UE}(\bm{s}^-_\mathrm{UE}[n]) \in \mathbb{R}^{6 \times 6}$
and $\bm{q}_\mathrm{UE}(\bm{s}^-_\mathrm{UE}[n])\in\mathbb{R}^{6}$
denote the observed \gls{FIM} and gradient of the log-likelihood function of \gls{UE} position given \gls{DoD} estimates from multiple \glspl{BS}.

Let $\bm{m}\in\mathbb{R}^{2K}$ denote the estimated \glspl{DoD} of $K$ \glspl{BS} towards a \gls{UE}. It follows from \eqref{eq:dod_dist} that $\bm{m}\sim\mathcal{N}\left(\bm{\mu}(\bm{p}),\bm{C}\right)$, where
\begin{align}
    \bm{\mu}(\bm{p}) & = \left[\vartheta_1(\bm{p}),\varphi_1(\bm{p}),\ldots,\vartheta_K(\bm{p}),\varphi_K(\bm{p})\right]^{\textrm{T}} \\
    \bm{C} & = \mathrm{blkdiag}\left\{\bm{C}_1,\ldots,\bm{C}_K\right\}.
\end{align}
Here, $\bm{p}\in \mathbb{R}^{3}$ denotes the $3$D Cartesian coordinate of a \gls{UE}'s position and $\mathrm{blkdiag}\{\cdot\}$ denotes a block-diagonal matrix. Note that 
\begin{align}
    \vartheta_k(\bm{p}) & =\arctan\left(\frac{-\Delta z_k}{d_{{2D}_k}}\right)+\pi/2\\
    \varphi_k(\bm{p}) & =\arctan2\left(\Delta y_k,\Delta x_k\right),
\end{align}
where $d_{{2D}_k} = \sqrt{\Delta x^2_k+\Delta y^2_k}$, $\Delta x_k=x_\mathrm{UE}-x_{\mathrm{BS}_k}$, $\Delta y_k=y_\mathrm{UE}-y_{\mathrm{BS}_k}$, and $\Delta z_k=z_\mathrm{UE}-z_{\mathrm{BS}_k}$. The gradient of the log-likelihood function of $\bm{p}$ given $\bm{m}$, and respective observed \gls{FIM}, now follow from \cite[Ch.3]{Kay93}
\begin{align} \label{eq:grad_fim_pos}
    [\bm{q}_\mathrm{UE}(\bm{p})]_m & = \left(\frac{\partial \bm{\mu}(\bm{p})}{\partial [\bm{p}]_m}\right)^{\textrm{T}} \bm{C}^{-1}\left(\bm{m} - \bm{\mu}(\bm{p})\right) \\
    [\bm{\mathcal{I}}_\mathrm{UE}(\bm{p})]_{m,n} & \approx  \left(\frac{\partial \bm{\mu}(\bm{p})}{\partial [\bm{p}]_m}\right)^{\textrm{T}} \bm{C}^{-1} \frac{\partial \bm{\mu}(\bm{p})}{\partial [\bm{p}]_n}.
\end{align}
\smallskip

%% file: simulations.tex
\section{Numerical Results} \label{sec:simulations}

\subsection{Deployment Scenario}

\begin{figure}[!t]
\centerline{\includegraphics[width=\linewidth]{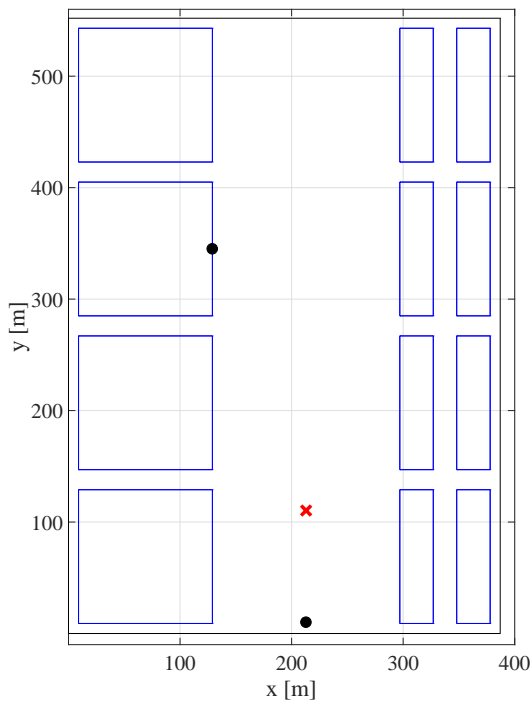}}
\caption{Illustration of the deployment scenario considered in this paper for assessing the performance of the proposed $2$-stage \gls{EKF}. We have modified the Madrid grid in order to have a larger open area between \glspl{BS} (black dots) and \gls{UE} (red cross). The radio channel between \gls{UE} and \glspl{BS} is according to the METIS ray-tracing channel model \cite{metis_channel_2015}.}
\label{fig:deployment}
\end{figure}

We consider a scenario where two \glspl{BS} and a \gls{UE} are deployed on the Madrid grid. In particular, a modification to the original Madrid grid is considered in this paper in order to obtain a large (up to \SI{500}{\meter} long) open area. This is a common envisioned deployment for mmW cellular systems. Fig.~\ref{fig:deployment} illustrates the modified Madrid grid considered in here as well as the locations of the \glspl{BS} and \gls{UE}. In particular, \glspl{BS} are deployed at a height of \SI{50}{\meter} while that of the \gls{UE} is \SI{1.5}{\meter}.

The mmW system considered in this numerical study operates at \SI{39}{\giga\hertz} with a bandwidth of \SI{200}{\mega\hertz} and subcarrier spacing of \SI{120}{\kilo\hertz}. The number of subcarriers available for transmitting \gls{DL}-\glspl{RS} is $1656$. The power budget at each \gls{BS} is \SI{21}{\decibel m}. Each \gls{BS} transmits \gls{DL}-\glspl{RS} through $64$ beams pointing in different directions. In particular, such \gls{BS} beams span $40^\circ$ both in elevation and azimuth, and the \SI{3}{\decibel} beamwidth is $\approx3^\circ$. \gls{DL}-\glspl{RS} for different beams and \glspl{BS} are assumed to be scheduled in orthogonal radio resources. The \gls{UE} receives \gls{DL}-\glspl{RS} from $52$ beams spanning $360^\circ$ in azimuth and a fixed direction ($\approx75^\circ$) in co-elevation. The \SI{3}{\decibel} beamwidth is $\approx6^\circ$ in azimuth and $\approx40^\circ$ in elevation. The maximum gains of the \gls{BS} and \gls{UE} beams are \SI{\approx 30}{\decibel i} and \SI{\approx17}{\decibel i}, respectively. The \gls{UE} measures the \gls{BRSRP} for all $64\times52$ beam-pairs, for both \glspl{BS}, in \SI{160}{\milli\second}, after which it feedbacks the highest \glspl{BRSRP}. The amount of feedback \glspl{BRSRP} may be signalled by the network, for example.

The radio channels between a given \gls{UE} and \glspl{BS} are modelled according to the METIS ray-tracing channel model \cite{metis_channel_2015}. Hence, all multipath components between the \gls{UE} and \glspl{BS} are taken into account in the \glspl{BRSRP} measurements, and re-calculated for every \gls{UE} position.

\subsection{Performance of the Proposed \gls{EKF}}

We assess the performance of the proposed two-stage \gls{EKF} by considering a \gls{UE} moving with a velocity of \SI{2}{\meter\per\second}. The \gls{UE} moves along a \SI{100}{\meter}-long straight trajectory with a south-to-north direction. The starting position of the \gls{UE} is illustrated in Fig.~\ref{fig:deployment}. The southernmost \gls{BS} is north-facing while the northernmost \gls{BS} has a \SI{60}{\degree} orientation clockwise from East-side. The \gls{UE} feedbacks \gls{BRSRP} measurements every \SI{160}{\milli\second}. Initialization of the \gls{DoD}-\gls{EKF} and Pos-\gls{EKF} follows that in \cite{KCWH17}.

In particular, we consider the case when the number of feedback \gls{BRSRP} measurements (corresponding to beam-pairs) is $16$, $8$, $6$, and $5$. For less than five \gls{BRSRP} measurements the corresponding angle-domain ambiguity function is far from the ideal Dirac-delta, and the resulting likelihood function has multiple global maxima. Hence, the performance of the proposed \gls{EKF} degrades rapidly when the number of feedback \gls{BRSRP} measurements is smaller than five. We emphasize that the performance of the proposed \gls{EKF} with respect to the number of feedback \gls{BRSRP} measurements is heavily dependent on the shape of the \glspl{BS}' transmit beams. Therefore, one can achieve sub-meter $3$D positioning accuracy with, say, three \gls{BRSRP} measurements given that the \glspl{BS}' transmit beams have the necessary characteristics in terms of angle-domain ambiguity function.

\begin{figure}[t]
\centerline{\includegraphics[width=1.15\linewidth]{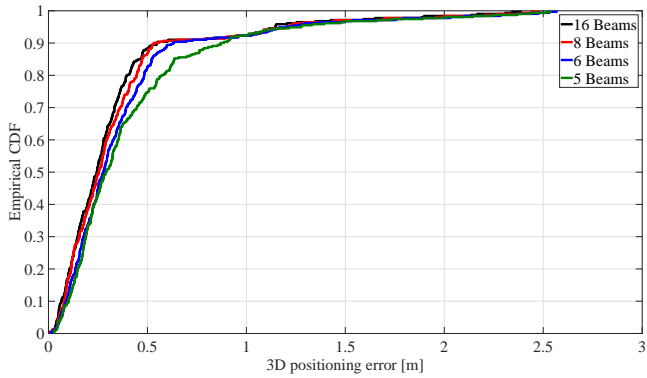}}
\caption{Empirical \gls{CDF} of $3$D positioning error obtained using the proposed \gls{EKF} for a varying number of feedback \gls{BRSRP} measurements. Reporting the five beams (out of the available $64$ beams) corresponding to the highest \gls{BRSRP} measurements suffices in achieving sub-meter $3$D positioning accuracy in \SI{90}{\percent} of the \gls{UE}'s trajectory. This is due to the high directivity of the employed transmit beams and allows one to optimize the capacity of the feedback channel.}
\label{fig:cdf_pos}
\end{figure}

\begin{figure}[t]
\centerline{\includegraphics[width=1.15\linewidth]{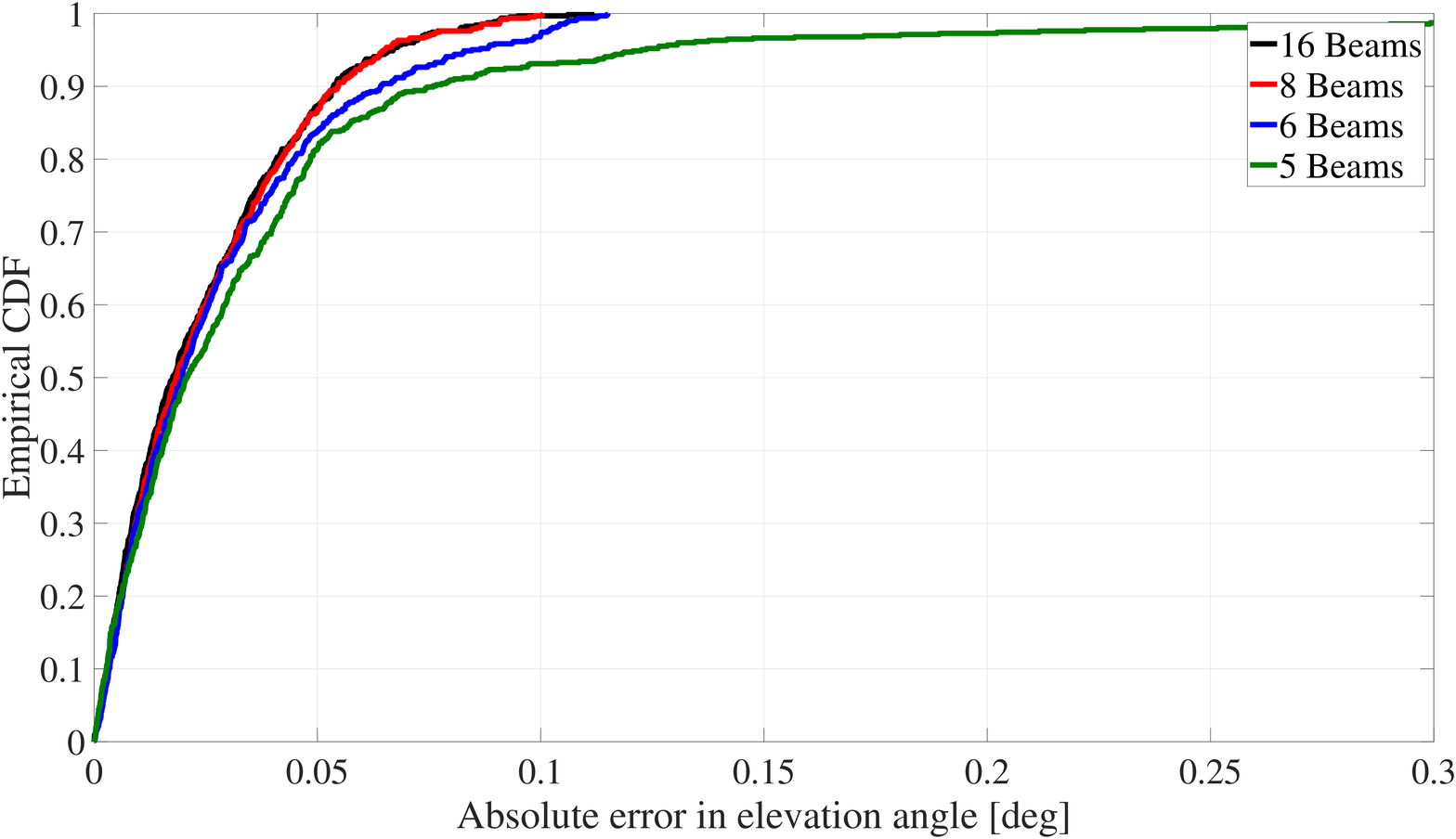}}
\caption{Empirical \gls{CDF} of elevation-angle error obtained using the proposed \gls{EKF} for a varying number of feedback \gls{BRSRP} measurements. Increasing the number of feedback beams improves the accuracy only slightly. This is due to the high directivity of the employed transmit beams.}
\label{fig:cdf_el}
\end{figure}

\begin{figure}[t]
\centerline{\includegraphics[width=1.15\linewidth]{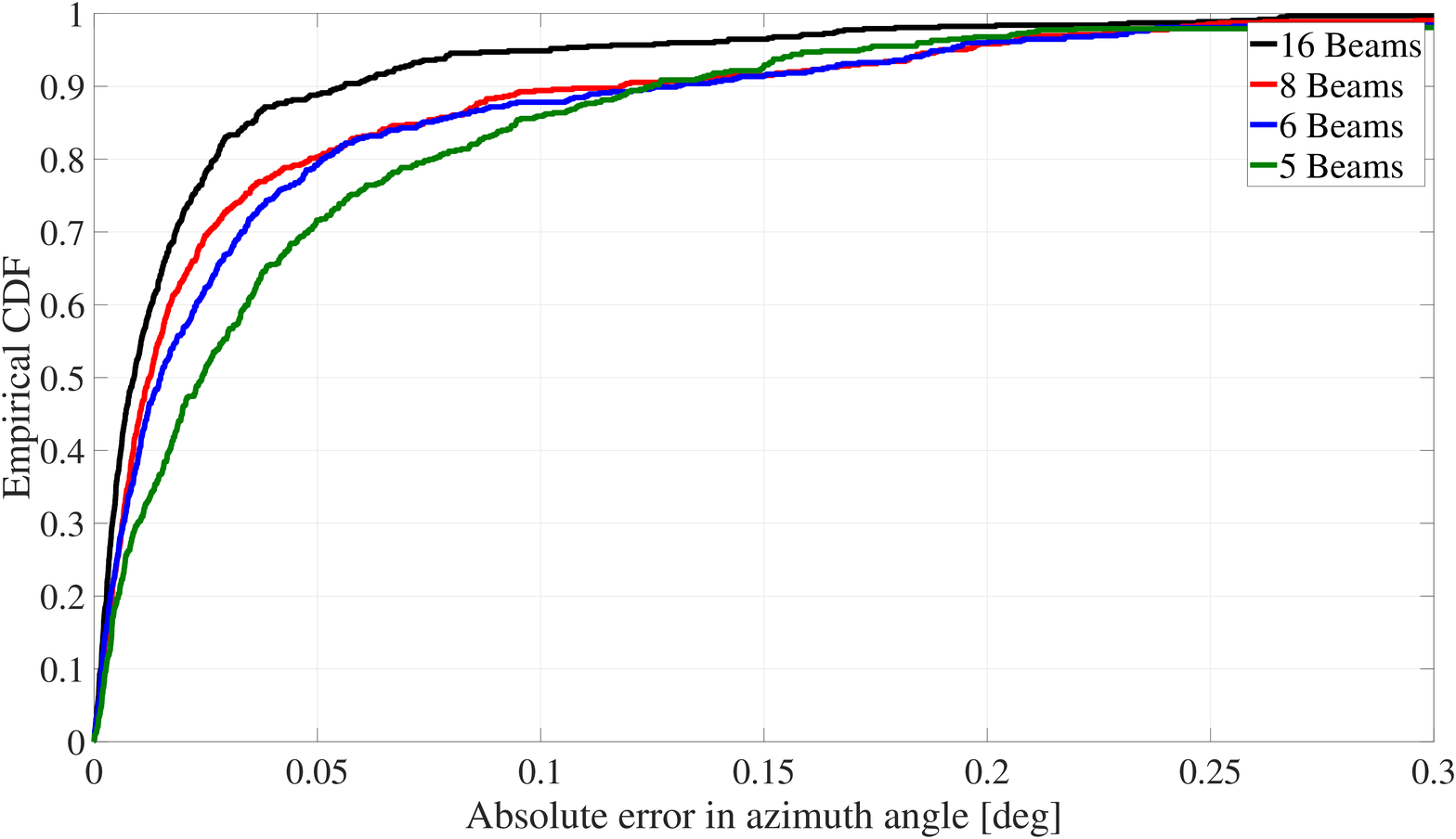}}
\caption{Empirical \gls{CDF} of azimuth-angle error obtained using the proposed \gls{EKF} for a varying number of feedback \gls{BRSRP} measurements. Increasing the number of feedback beams improves the accuracy only slightly. This is due to the high directivity of the employed transmit beams.}
\label{fig:cdf_az}
\end{figure}

\begin{figure}[t]
\centerline{\includegraphics[width=1.15\linewidth]{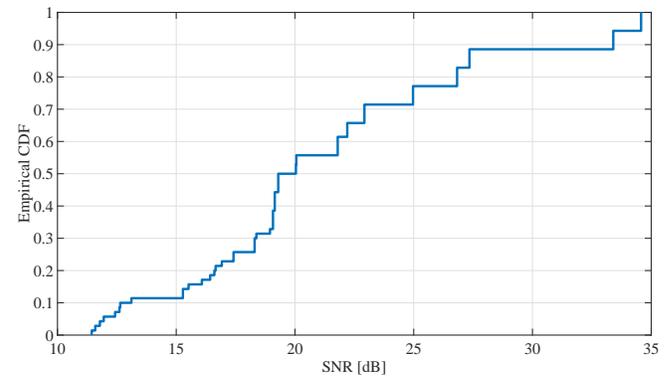}}
\caption{Empirical \gls{CDF} of received \gls{SNR} at the \gls{UE}, after gain from transmit/receive beams, for the $16$ beams considered for reporting.}
\label{fig:cdf_snr}
\end{figure}

The performance metrics employed to assess the performance of the proposed \gls{EKF} are the 3D positioning error $(\sqrt{\tilde{x}^2 + \tilde{y}^2 + \tilde{z}^2})$, elevation angle error $(|\tilde{\vartheta}|)$ as well as azimuth angle error $(|\tilde{\varphi}|)$. The corresponding \glspl{CDF} are illustrated in Figs.~\ref{fig:cdf_pos}-\ref{fig:cdf_az}. The \gls{CDF} of the received \gls{SNR} per beam is also given in Fig.~\ref{fig:cdf_snr}. Results show that reporting the five beams (out of the available $64$ beams) corresponding to the highest \gls{BRSRP} measurements suffices in achieving sub-meter $3$D positioning accuracy in \SI{90}{\percent} of the \gls{UE}'s trajectory. Also, increasing the number of feedback beams improves the positioning accuracy only slightly. This may be understood by the high directivity of the employed transmit beams. In particular, reporting \gls{RSRP} measurements of beams that have a main-lobe towards directions away from the \gls{LoS} between \gls{BS} and \gls{UE} leads to a marginal increase (and may even be detrimental for low-SNR) in angle-related information compared to the beams that point towards the \gls{UE}. In practice, this is important since it allows one to optimize the capacity of the feedback channel.

%% file: conclusion.tex
\section{Conclusion} \label{sec:conclusions}

In this article, we proposed a 3D \gls{UE} positioning method for \gls{5G} \gls{mmW} networks by exploiting $3$D downlink beamforming from base-stations. More specifically, the proposed sequential 3D \gls{UE} position estimation was performed at the network-side by means of a two-stage \gls{EKF} and based on feedback beam-\gls{RSRP} measurements carried out at the \gls{UE}. In particular, in the first \gls{EKF} stage, the directions-of-departure of the beamformed \gls{DL} \glspl{RS} in the feedback scheme were estimated and tracked at \glspl{BS}, whereas in the second angle-based \gls{EKF} stage, such angle estimates were fused from all available \glspl{BS} into 3D position estimates at a central entity. Performance results of the proposed algorithm on a realistic outdoor \gls{5G} deployment based on the METIS ray-tracing propagation model show that sub-meter $3$D positioning accuracy of users is achievable in 90\% of the cases by reporting only $8$ base-station DL beams.

Future work includes taking into account uncertainties in \glspl{BS}' orientation for \gls{UE} positioning.